\documentclass[aps,prl,superscriptaddress,floatfix,reprint]{revtex4-2}
\usepackage{graphicx}
\usepackage{amsmath}
\usepackage{amssymb}
\usepackage{amsthm}
\usepackage{hyperref}
\usepackage{color}
\usepackage{CJKutf8}
\usepackage{bbm}
\usepackage{bbold}
\usepackage{tikz}

\newcommand{\bmat}{\left ( \begin{array}{cc}}
\newcommand{\emat}{ \end{array}\right )}

\renewcommand\epsilon\varepsilon
\renewcommand\phi\varphi

\newcommand\be{\begin{eqnarray}}
\newcommand\ee{\end{eqnarray}}
\newcommand{\nn}{\nonumber}

\newcommand{\vev}[1]{\left\langle #1\right\rangle}
\newcommand{\Tr}{{\text{Tr}}}

\newcounter{yjcc}

\begin{document}

\title{Parisi's Hypercube, Fock-Space Frustration and Near-AdS$_2$/Near-CFT$_1$ Holography}
\author{Micha Berkooz}
\email{micha.berkooz@weizmann.ac.il}
\affiliation{Department of Particle Physics and Astrophysics, Weizmann Institute of Science, Rehovot 7610001, Israel}

\author{Yiyang Jia\begin{CJK*}{UTF8}{gbsn}
		(贾抑扬)
\end{CJK*}}
\email{yiyang.jia@weizmann.ac.il}
\affiliation{Department of Particle Physics and Astrophysics, Weizmann Institute of Science, Rehovot 7610001, Israel}

\author{Navot Silberstein}
\email{navotsil@gmail.com}
\affiliation{Department of Particle Physics and Astrophysics, Weizmann Institute of Science, Rehovot 7610001, Israel}

\date{\today}

\begin{abstract}
 We consider a model of Parisi where a single particle hops on an infinite-dimensional hypercube, under the influence of a uniform but disordered magnetic flux.  We reinterpret the hypercube as the Fock-space graph of a many-body Hamiltonian  and the flux as a frustration of the return amplitudes in Fock space. We will identify the set of observables that have the same correlation functions as the double-scaled Sachdev-Ye-Kitaev (DS-SYK) model, and hence the hypercube model is an equally good quantum model for near-AdS$_2$/near-CFT$_{1}$ (NAdS$_2$/NCFT$_1$) holography. Unlike the SYK model, the hypercube Hamiltonian is not $p$ local. Instead, the SYK model can be understood as a Fock-space model with similar frustrations. Hence we propose this type of Fock-space frustration as the broader characterization for NAdS$_2$/NCFT$_1$ microscopics, which encompasses the hypercube and the DS-SYK models as two specific examples. We then speculate on the possible origin of such frustrations.
\end{abstract}
\maketitle
\allowdisplaybreaks[3]
Two-dimensional nearly anti-de Sitter (NAdS$_2$) spacetime arises ubiquitously as the near-horizon geometry of near-extremal black holes in higher dimensions. In holographic theories, this means that in the appropriate near-extremal states an AdS$_{D+1}$/CFT$_{D}$ ($D > 1$) duality flows to a near-AdS$_{2}$/near-CFT$_{1}$ (NAdS$_2$/NCFT$_1$) duality at low energy. In fact, considerable progress has been made by directly constructing microscopic models for NCFT$_{1}$ (nearly conformal field theory in one dimension), the most notable of which is the Sachdev-Ye-Kitaev (SYK) model \cite{kitaev2015, maldacena2016, sachdev1993, french1970, bohigas1971}. In the SYK model, a system of $N$ Majorana fermions interact through a $p$-body interaction in which a fermion can couple to any of the rest. At low energy, the model's thermodynamics and correlators reproduce those of the Jackiw-Teitelboim gravity---a dilaton gravity theory that can arise by dimensionally reducing higher-dimensional gravity to NAdS$_2$ spacetime \cite{maldacena2016a}. In addition, the SYK model is also important as a solvable model of quantum chaos in $p$-local systems \textit{per se} (and can probably also be realized experimentally). 
Its low-energy solution is obtained by using Schwinger-Dyson equations in the limit  $N\to \infty$ with $p$ fixed. But the double-scaled SYK (DS-SYK) limit  $p, N\to\infty$ with $ \lambda=2p^2/N$ fixed  can be solved exactly in $\lambda$  for all energy scales using combinatorics. The latter technique also allows for the reconstruction of the AdS$_2$ dynamics (and generalizes it to a $q$-deformed AdS$_2$) \cite{Berkooz:2018qkz,Berkooz:2018jqr,lin2022, berkooz2022quantum}.

There are, however, additional models that are not even  $p$ local but have the same combinatorial solution. The simplest such example is the hypercube model of Parisi \cite{Parisi:1994jg}, made out of $d$ qubits, along with a Hamiltonian with interactions that couple together all degrees of freedom in each term. It is therefore interesting to identify which microscopic aspects of the SYK model are essential and which are spurious for the NAdS$_2$/NCFT$_1$ holography, as well as clarify whether quantum chaos is similar in these models.  We will  try to pinpoint exactly what the two models have in common, in terms of dynamics and in terms of the appropriate set of observables, and use it as a stepping stone toward a broader characterization of  NAdS$_2$/NCFT$_1$ microscopics. We can then hope it is this broader characterization that survives the examination from the AdS$_{D+1}$/CFT$_{D}$ ($D > 1$) viewpoint. More details are covered in a companion article \cite{berkooz2023parisis}.

Parisi introduced a $d$-dimensional hypercubic model where there are superconducting dots living on the hypercube vertices, whose energy is frustrated by a uniform (position-independent) but disordered magnetic flux.  Here we remove the superconducting dots of the original theory, and the physics becomes that of a single particle hopping on the hypercube vertices under the influence of the same flux.
By doing so we drastically change the role of the flux: the flux frustrates energies in the original theory but now frustrates the return amplitudes of a hopping particle. The former tends to increase the glassiness of a system, and the latter tends to delocalize wave functions and hence to thermalize a system. Nevertheless, Parisi's key insight that such high-dimensional fluxed operators coarse grain to a $q$-deformed oscillator, remains. Its similarity with the DS-SYK model was first noticed by \cite{garciaprivate2019,Jia_2020}, and given the simplicity of model,  one cannot help but wonder if this similarity extends to correlation functions.  We will answer this question in the affirmative.

The hypercube has $2^d$ vertices which we denote by $\{-1/2,+1/2\}^d$ with $\sigma^3_\mu/2\ (\mu=1,\ldots,d)$ being the position operators of the particle. We will use a gauge that is different from Parisi's original choice.  The point is to use a rotationally covariant gauge so that the insertions of probe operators become much simpler.  Our Hamiltonian is
\begin{equation}\label{eqn:symmetricGauge}
    H=- \frac{1}{\sqrt{d}}\sum_{\mu=1}^d  D_\mu:=-\frac{1}{\sqrt{d}}\sum_{\mu=1}^d ( T_\mu^++ T_\mu^-),
\end{equation}
where  $T^-_\mu=(  T^+_\mu)^\dagger$  and
 \begin{equation}\label{eqn:hoppingDef}
       T^+_\mu = \prod_{\nu=1, \nu\not=\mu}^d e^{\frac i4 F_{\mu\nu}\sigma^3_\nu} \sigma^+_\mu,\quad \sigma^+_\mu =\frac{\sigma^1_\mu + i \sigma^2_\mu}{2}.
\end{equation}
The $\sigma_\mu^i (i=1, 2, 3)$  is the $i$th Pauli matrix acting on the $\mu$th qubit, and $F_{\mu\nu}$ is the antisymmetric tensor of the  background flux.  We have chosen a normalization for  $H$ such that it  has a compact spectral support at  $d = \infty$ (as we will also do for the DS-SYK model).  The fluxes  $F_{\mu\nu}$ are quench disordered,  identically and independently distributed  with the additional requirement that the distribution is even so that ($\langle \cdot \cdot \rangle$ stands for an ensemble average)
\begin{equation}
    \vev{\sin F_{\mu\nu}} =0.
\end{equation}
The distribution is otherwise completely general and 
\begin{equation}
   q:=\vev{\cos F_{\mu\nu}} 
\end{equation}
is a tunable parameter.  

$T^+_\mu$ is a hopping operator that transports the particle in the forward $\mu$ direction, while assigning to it a random phase due to the disordered flux. The holonomy  $T^-_\nu  T^-_\mu  T^+_\nu  T^+_\mu$ in the $\mu\nu$ plane then gives the return amplitude of hopping counterclockwise around this plaquette. However, it is more convenient to study the holonomy in terms of $D_\mu$ operators, which combines the forward and backward hoppings,
\begin{equation}\label{eq:frust}
\begin{split}
     & {\cal W}_{\mu\nu}= D_\nu D_\mu D_\nu D_\mu=\cos F_{\mu\nu} - i \sin F_{\mu\nu} \sigma^3_\mu  \sigma^3_\nu,\\
     & \langle {\cal W}_{\mu\nu} \rangle = q.
\end{split}
\end{equation}
We can also think of ${\cal W}_{\mu\nu}$ as the mutual frustration of different terms in the Hamiltonian. 

We can view the Hamiltonian \eqref{eqn:symmetricGauge} as a many-body system of $d$ interacting qubits with the hypercube being its Fock-space graph \cite{altshuler1997}: if we view each basis vector as a point,  and connect two points whenever the corresponding basis vectors have a nonzero transition amplitude, then we get back  to the picture of a single particle hopping on a hypercube. The many-bodiedness is encoded in the requirement that a Fock-space graph should have a diverging vertex degree ($d\to \infty$).   In this manner, we have reinterpreted the hypercube as living in a Hilbert space rather than real space. 

\begin{figure}
  \begin{center}
\begin{tikzpicture}[scale=0.7]
%% vertices
\draw[thick] (0,0) circle (35pt);

\filldraw (-0.72,-1) circle (1pt);
 \node at (-0.89,-1.2) {$\mu$};
 \filldraw (0.72,1) circle (1pt);
 \node at (0.89,1.2) {$\mu$};
 \draw[] (-0.72,-1) -- (0.72,1);

 \filldraw (-1.18, 0.4) circle (1pt);
 \node at (-1.4, 0.4) {$\nu$};
 \filldraw (1.18, 0.4) circle (1pt);
 \node at (1.4, 0.4) {$\nu$};
  \draw[](-1.18, 0.4) -- (1.18, 0.4) ;

 \filldraw (-1.18, -0.4) circle (1pt);
 \node at (-1.4, -0.4) {$\rho$};
 \filldraw (1.18, -0.4) circle (1pt);
 \node at (1.4, -0.4) {$\rho$};
  \draw[](-1.18, -0.4) -- (1.18, -0.4) ;

%% vertices
\draw[thick] (5,0) circle (35pt);

\filldraw (4.28,-1) circle (1pt);
\filldraw (5.4,1.17) circle (1pt);
\draw[dashed] (4.28,-1) -- (5.4,1.17);

\filldraw (3.86, -0.5) circle (1pt);
\filldraw (6.14, -0.5) circle (1pt);
\draw[](3.86, -0.5) -- (6.14, -0.5);

\filldraw (5.6,-1.07) circle (1pt);
\filldraw (4.4, 1.07) circle (1pt);
\draw[] (5.6,-1.07) -- (4.4, 1.07);

\end{tikzpicture}

\end{center}
\caption{Left: a chord diagram contributing to $2^{-d}\vev{\Tr H^6}$, which represents the hopping sequence $D_\nu D_\rho D _\mu D_\rho D_\nu D_\mu$.  This diagram has a value of $q^2$. Right: a chord diagram contributing to a two-point insertion $2^{-d}\vev{\Tr H^2 O H^2 O}$. The dashed line represents the $O$ chord and the solid lines represent the $H$ chords.  This diagram has a value of $q \tilde q^2$.} \label{fig:chordExamples}
\end{figure}
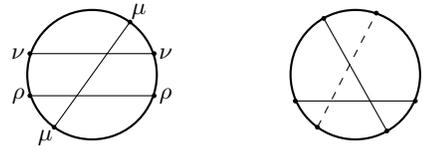

The spectrum of the model is solved by moment method via $q$-deformed oscillators \cite{Parisi:1994jg,Marinari:1995jwr,Cappelli-1998}.  The $2k$th moment of the hypercube model can be written as 
\begin{align}\label{eqn:moments}
   2^{-d}\vev{ \Tr H^{2k}}=&2^{-d}d^{-k} \sum_{\{\mu_i\}}  \langle \Tr
  \   D_{\mu_1}  D_{\mu_2}\ \ldots  D_{\mu_{2k}}
    \rangle.
\end{align}
Since the trace is a sum of return amplitudes, a forward hopping must be paired with a backward hopping, which means the subscripts $\mu_1, \ldots, \mu_{2k}$  must form $k$ pairs.  At leading order in  $1/d$, we can focus on the case where these $k$  indices are all distinct (any further coincidence among the $k$ pairs will be suppressed by $1/d$).  We can use chord diagrams to represent such pairings: draw $2k$ points on a circle representing the subscripts, and connect two points by a chord if the corresponding subscripts are paired.  We illustrate one example in the left panel of Fig. \ref{fig:chordExamples}. 
To evaluate a chord diagram, we can move the operators until the paired operators become adjacent to each other, and in the process we generate phase terms by applying Eq. \eqref{eq:frust}  repeatedly (and that $D_\mu^2=1$).  
% Note that the $\sin F$ term  cannot contribute to moments  at leading order since we required $\vev{\sin F}=0$, and $\vev{\sin^2 F}$ will not appear until $1/d^2$ order. 
The result is that we pick up an independent $\cos F$ for each interlacing ordering of two pairs of hoppings and the moments are 
\begin{equation}\label{eqn:HchordRules}
     2^{-d}\vev{ \Tr H^{2k}}=\sum_{\text{diagrams}} q^\text{number of chord intersections}.
\end{equation}
The corresponding spectral density is given by \cite{ismail1987} (an efficient way of evaluating the sum using a transfer matrix is given in \cite{Berkooz:2018qkz}):
\begin{align}\label{eq:RhoChords} 
  &  \rho(E)= \frac{\Gamma_{q^2}\left(\frac{1}{2}\right)}{\pi \sqrt{1+q}} \left[1- \frac{E^2}{4}(1-q)\right]^\frac{1}{2} \prod_{l=1}^\infty \left[1- \frac{(1-q)q^l E^2}{(1+q^l)^2}\right],\nn
\\
   & \Gamma_{q^2}\left(\frac{1}{2}\right)= \sqrt{1-q^2} \prod_{j=0}^\infty (1-q^{2j+2}) (1-q^{2j+1})^{-1}.
\end{align}
The double-scaled SYK model can be solved in a similar way \cite{erdos2014, cotler2016} and the coincidence with the hypercube model's spectral density was noted in \cite{garciaprivate2019,Jia_2020}.  In this Letter we will extend the similarity to correlation functions and explain why the coincidence is not accidental at all.   

The SYK model is as follows. Consider $N$ Majorana fermions $\{\psi_{i},\psi_{j}\} =2\delta_{ij},\ i,j=1,\ldots,N$ and the Hamiltonian
\begin{align}\label{eqn:SYKhamiDef}
       H_\text{SYK}= \Sigma_I J_I \Psi_I
\end{align}
where $I$ is a multi-index of length $p$ ($p$ is an even integer),
\begin{equation}
\begin{split}
    &I = \{i_1, i_2, \ldots, i_p\}, \quad  1\leq i_1<i_2<\cdots<i_p \leq N,\\
    &\Psi_I = i^{p/2} \psi_{i_1}\psi_{i_2} \cdots \psi_{i_p},\ \ (\Psi_I^2=1) \ .
\end{split}
\end{equation}
 Moreover, $J_I$ are Gaussian random variables that are independently and identically distributed with variance
 \begin{equation}
     \vev{J_I^2} = \binom{N}{p}^{-1}.
 \end{equation}
 
The main feature that the SYK model shares with the Parisi model is a similar structure of holonomies encoding frustrations, given by 
 \begin{equation}\label{eq:SYKFrust}
    {\cal W}_{IJ}= \Psi_I\Psi_J\Psi_I\Psi_J = (-1)^{|I\cap J|},
\end{equation}
where $|I\cap J|$ is the cardinality of the intersection of $I$ and $J$. The subscript $I$ plays a similar role as $\mu$ does in the hypercube model (which specifies there the direction of hopping). Comparing with Eq. \eqref{eq:frust}, we see that the SYK model frustrations are generated by uniform fluxes of $0$ and $\pi$. By uniformity we mean that the holonomy ${\cal W}_{IJ}$ only depends on $I$ and $J$, but does not depend on which state it acts on. Namely, in the Fock space a general loop  produces a phase that depends on its shape and orientation, but is independent of its position.  To accomplish a complete analogy with the hypercube model, we would still need the holonomies on different plaquettes to be statistically 
independent and have a tunable average value.  This is achieved by going to the double-scaled SYK limit:
 \begin{equation}
     N,p\rightarrow \infty,\  \text{with fixed }\frac{p^2}{N}.
 \end{equation}
In this limit multi-index intersections become an independently random process  for each pair of $I$ and $J$, and $|I\cap J|$ is Poisson distributed with a mean value $p^2/N$,  giving an average holonomy 
\begin{equation}
    q= \langle (-1)^{|I\cap J|} \rangle_{I,J} = e^{-2p^2/N},
\end{equation}
where the average is over all possible values of $I$ and $J$ \cite{erdos2014}. This $q$ plays the identical role in the DS-SYK model as the $\vev{\cos F}$  plays in the hypercube model.

{\bf $p$-local vs. frustrated Hamiltonians:}
The SYK Hamiltonian \eqref{eqn:SYKhamiDef} is manifestly $p$ local ($p$ being the length of the interaction). The Parisi Hamiltonian is not of that form, as each term in \eqref{eqn:hoppingDef} depends on all the available qubits. Nevertheless, the solution is the same. The real criterion that allows for the same solution using chord diagrams is the fact that the frustrations satisfy 
\begin{equation}\label{eq:LowFrust}
    [{\cal W}_{\mu\nu}, D_\rho]=0\ \text{or}\  [{\cal W}_{IJ}, \Psi_K]=0
\end{equation}
with probability 1 in the thermodynamic limit. The holonomies, or frustrations, are effectively short ranged and do not interfere with most of the many-body interaction terms. This is another way of phrasing the uniformity requirement for the frustrations.

{\bf Observables:} To exhibit a full solution of the Parisi model  at the same level as the DS-SYK model, we need a rich enough set of observables and show that their correlation functions are the same. As we shall see below, the chord combinatorics for probes in the hypercube model is again identical to that of the DS-SYK model. As a consequence they develop the same infrared behavior, which implies the hypercube model also has a NCFT$_{1}$ limit and its out-of-time-order correlator  has an exponential growth in time with a maximal Lyapunov exponent (which matches the fast-scrambling nature of black holes \cite{sekino2008, shenker2014, maldacena2015}). We shall see that the operator conformal dimensions in both models can be understood as a ratio of frustrations.

What are the appropriate probe operators in this model? Consider how we probe a near-extremal black hole in higher-dimensional AdS$_{D+1}$. We expect that single-trace operators of the dual higher-dimensional CFT$_D$  become complicated by the time they flow to NCFT$_1$. Therefore, our best chance is to give a statistical description for them.  The Hamiltonian is one of the single-trace operators, so we may expect other single-trace operators to have a similar form. Since the Hamiltonian \eqref{eqn:symmetricGauge} is built from hopping operators $D_\mu$, we  suggest the following class of operators as suitable observables:
\begin{equation}\label{eqn:parisiProbe}
    O = -\frac{1}{\sqrt{d}}\sum_{\mu=1}^d \tilde D_\mu :=  -\frac{1}{\sqrt{d}}\sum_{\mu=1}^d (\tilde T_\mu^+  +  \tilde T_\mu^-),
\end{equation}
where $\tilde T_\mu^+$ is defined in the same manner as $ T_\mu^+$ in Eq. \eqref{eqn:hoppingDef}, but with a different uniform and disordered flux $\tilde F_{\mu\nu}$ which may or may not  correlate with $F_{\mu\nu}$. 
Similar logic applies to SYK probes and suggests that they can be chosen to be a product of $\tilde p$ fermions
$ O_{\text{SYK}}=    \sum_{\tilde{I}} \tilde{J}_{\tilde{I}} \Psi_{\tilde{I}}
$,
where $\tilde I$ is an index set of length $\tilde p$ \cite{Berkooz:2018qkz,Berkooz:2018jqr}. 
We can generalize Eq. \eqref{eqn:parisiProbe} further and take $O$ to be a sum of products of a finite number of hoppings, twisted by random phases, but this does not add any new physics as far as observables are concerned. It does open new options for Fock-space dynamics, as we will discuss later.

An odd number of insertions of $\tilde D_\mu$ are exponentially suppressed because $\langle {D \tilde D} \rangle = \langle\cos [( F- \tilde F)/4] \rangle^{d-1} \to 0$,  and hence we only consider an even number of insertions. Moments with two-point insertions have the form
\begin{align}\label{eqn:twoPtTrace}
   \vev{ \Tr H^{k_2}OH^{k_1}O}=&\frac{1}{d^{\frac{k_1+k_2+2}{2}}} \sum_{\nu_1,\nu_2,\{\mu_i\}}  \langle \Tr
  \   D_{\mu_1}\ldots  D_{\mu_{k_2}}\nn \\
&\quad \tilde D_{\nu_1} D_{\mu_{k_2+1}} \ldots  D_{\mu_{k_1+k_2}}
   \tilde D_{\nu_2}
    \rangle.
\end{align}
Because of the same exponential suppression,  two $\tilde D$'s must pair up and the $D$'s must pair up among themselves. Therefore, we can obtain the two-point functions by chord diagrams where one type of chord (marked by dashed lines) connects the $O$ insertions ($O$ chords), and another type connects the Hamiltonians ($H$ chords). We draw an example in the right panel of Fig. \ref{fig:chordExamples}.  Note also that
\begin{align} \label{eqn:DtildeDfrust}
&\tilde D_\nu D_\mu \tilde D_\nu D_\mu =\cos \frac{F_{\mu\nu}+ \tilde F_{\mu\nu}}{2}  - i \sin \frac{F_{\mu\nu}+ \tilde F_{\mu\nu}}{2}\sigma^3_\mu  \sigma^3_\nu, \nn\\
&\vev{\tilde D_\nu D_\mu \tilde D_\nu D_\mu} := \tilde q = \vev{\cos \frac{F_{\mu\nu}+ \tilde F_{\mu\nu}}{2} },
\end{align}
which is a generalization of Eq. \eqref{eq:frust}. The remaining steps are entirely analogous to the discussion without insertions.
The two-point moment at leading order is  given by the sum of chord diagrams
\begin{align} \label{eqn:twoPtChordRules}
  &2^{-d}  \vev{ \Tr H^{k_2}OH^{k_1}O} \nn\\
  = &\sum_{\text{diagrams}}  q^{ \text{No. $H$-$H$ intersections}}\ \tilde q^{\text{No. $O$-$H$ intersections}},
\end{align}
and the four-point insertion rule works out similarly,
 \begin{align}\label{eqn:fourPtChordRules}
  &2^{-d}  \vev{ \Tr H^{k_4}OH^{k_3}O H^{k_2}OH^{k_1}O} \nn\\
  = &\sum_{\text{diagrams}}  q^{\# \text{$H$-$H$ intersections}}\ \tilde q^{ \text{No. $O$-$H$ intersections}} \nn\\
  &\hspace{3cm} \times \tilde q_{12}^{\text{No. $O$-$O$ intersections}},
\end{align}
where ``No. $H$-$H$ intersections'' means the total number of intersections among $H$ chords in a diagram and likewise for $O$-$H$ and $O$-$O$, where the weight for the latter is $\tilde q_{12}:= \langle\cos\tilde F_{\mu\nu}\rangle$.
These are exactly the same chord diagram rules for the DS-SYK model, and there the $q$ parameters are
\begin{equation}
    q = e^{-2p^2/N}, \quad \tilde q = e^{-2p\tilde p/N}, \quad \tilde q_{12} = e^{-2\tilde p^2/N}.
\end{equation}
The NCFT$_1$ limit of both models is given by \cite{Berkooz:2018jqr}
\begin{equation}
    q, \tilde q \to 1^-, \quad  \log \tilde q /\log q \ \  \text{fixed}
\end{equation}
in the temperature range 
\begin{equation}
  (-\log q)^{\frac{3}{2}}   \ll  T \ll (-\log q)^{\frac{1}{2}}.
\end{equation}
In this regime, the correlation functions have a conformal form and the operator dimensions are given by
\begin{equation}
    \Delta_O  = \log \tilde q /\log q, 
\end{equation}
which in the DS-SYK model implies $\Delta_O = \tilde p/p$ and in the hypercube model implies 
\begin{equation}\label{eqn:scalingDim}
      \Delta_O  =   \frac{\vev{(F_{\mu\nu}+\tilde F_{\mu\nu})^2}}{4 \vev{F_{\mu\nu}^2} }.
\end{equation}

{\bf Operator growth and the Parisi model as a typified SYK model:} Next we will argue that the Parisi model is a useful model for operator growth in the SYK model.
Consider first the ``growth" of $\rho=e^{-\beta H}$ as $\beta$ increases. This change can be encoded as evolution on \begin{equation}
    \text{span}\{  {\Psi}_{I_1}\ldots{\Psi}_{I_k},\ k\ge 0\}.
\end{equation}
Given that that $\Psi_I^2=1$, the evolution happens on the hypercube of operators
\begin{equation}
    d={N\choose p},\ (Z_2)^d\rightarrow   \left\{ \prod_I {\Psi}^{n_I}_I \Big| n_I\in\{0,1\}^d \right \}
\end{equation}
when we start the evolution at the origin. 
The right-hand side is an overcomplete set of operators but this is a valid description for motions that start at the origin and make fewer than ${\cal O}(N)$ hops (or we can go to the sparse SYK model \cite{swingle2020,garcia2021} where $d\sim N$ and alleviate the overcompleteness).

In slightly more detail, consider the plaquette whose corners are $\Psi_S \Psi_I^{0,1}\Psi_J^{0,1}$, where $\Psi_S$ includes all the other monomials' contribution. We can move from a corner by multiplying by $\Psi_K$ ($K= I$ or $J$) on the right. This is induced by ``evolution" in $\beta$. The flux on a plaquette is uniform in the sense before, and depends only on  $I \cap J$. To go to the Parisi model we now replace the overdetailed information of the phases by an average phase. So a typified version of operator growth dynamics in the DS-SYK model is just given by a Parisi model. 

For the growth of a more general operator it is important to choose the right class of operators. The most universal choice is to choose another random operator whose size scales as $\sqrt{N}$ (but with different coefficients than the Hamiltonian) and denote it by ${\cal O}_{\text{base}}$. Its Heisenberg time evolution now takes place in the hypercube,
 \begin{equation}
     (Z_2)^d\times(Z_2)^d\rightarrow   \{  \Pi_J {\Psi}^{m_J}_J {\cal O}_{\text{base}} \Pi_I {\Psi}^{n_I}_I |m_J,n_I\in\{0,1\}^d \}
\end{equation}
and it propagates the state both on the left and on the right lattice, i.e., the model is just the product of two lattices as above (but with a nontrivial inner product that mixes them, which is given by chord combinatorics). We can see that some standard measures of operator growth can be easily extracted from it. For example, evolution in the Krylov basis \cite{parker2019} is just the coarse information of the distance of the lattice point to the origin (for example in the way of \cite{Bhattacharjee2023, rabinovici2023bulk}). In the approach above, one can discuss the evolution of more complicated features of the operator by keeping more partial data about the location on the hypercube.

{\bf Fock space dynamics:}
Clearly the model can be generalized by including more complicated patterns of hoppings in Fock space. In fact, we can extrapolate between the pure Parisi model and SYK-type models as actions in Fock space by taking $H=\sum_{\alpha,A} W_{\alpha,A} O_{\alpha,A}$, where $O_{\alpha,A}$ is of the form
\begin{equation}
\begin{split}
    &\alpha=\{\mu_1,..\mu_p\}, A=\{n_1,..,n_p\},\ n_i=\pm,\\
    &\ O_{\alpha,A}= \Pi \sigma_{\mu_j}^{n_j}*\text{phase terms}.
\end{split}
\end{equation}
For example the complex SYK and complex DS-SYK models \cite{gu2019, berkooz2021} are precisely these models, with phase factors present in the Jordan-Wigner representation of fermions  and with the constraint that $\sum_i{n_i}=0$ to enforce the U(1) symmetry. This suggests some interesting generalizations of the SYK model relevant for physical situations. Consider a quantum dot of many fermions with a conserved charge. If the dot is tuned such that there are no $\psi^\dagger_i\psi_j$ terms in the Hamiltonian, then we expect that the model is given by a U(1)-invariant SYK model. But now we can generalize the U(1)-invariant model to
\begin{equation}
    H = \sum J_{ij}^{kl} \psi^i\psi^j\psi^{\dagger}_k\psi^{\dagger}_l e^{i\sum_m \phi_{ijkl,m} \psi^{\dagger}_m\psi^m}.
\end{equation}
The additional phases can all be small but there are many of them---as in the Parisi model, one cannot expand in the phases, but rather they can modify the infrared behavior. Another interesting application would be doing a similar construction using canonical bosons. The large amount of frustrations will make sure there is no low-temperature condensate \cite{jia2024bosonic}, which has been difficult to achieve in the $p$-local approach.

{\bf Discussions:} What is to be learned from such a picture? Minimally, we can say $p$-locality is not a broad enough characterization for NAdS$_2$/NCFT$_1$ microscopics. Indeed, $p$ locality describes a large class of models whose double-scaled limit gives NAdS$_2$/NCFT$_1$, an example other than the SYK model is the $p$-quantum-spin model where the double-scaled limit was first discovered \cite{erdos2014}. However, the hypercube model is not $p$ local, yet follows exactly the same combinatorics.  Instead, the Fock-space frustration picture encompasses both and hence is the broader characterization, which should be useful for model-building purposes. This is particularly important if we want to realize the NAdS$_2$/NCFT$_1$ relation as the infrared of a renormalization-group flow in a holographic CFT$_D$ ($D>1$) in an extremal black hole state---we should really be looking for signatures of frustrations rather than $p$ locality.

To summarize, we get  chord combinatorics [as in Eqs. \eqref{eqn:HchordRules}, \eqref{eqn:twoPtChordRules} and \eqref{eqn:fourPtChordRules}] and therefore automatically a NAdS$_2$/NCFT$_1$ duality  if a model has a Fock-space frustration which is 
\begin{enumerate}
    \item uniform and quench-disordered and
    \item independently and identically distributed on different (nonparallel) plaquettes of the Fock-space graph, with  a real and tunable average holonomy.
\end{enumerate}
The  NAdS$_2$/NCFT$_1$ emerges as the variance of the flux ($\vev{F_{\mu\nu}^2}$ in the Parisi model, $p^2/N$ in the DS-SYK model) is tuned to zero after the thermodynamic limit is taken. 

These criteria need to be understood as large-system-size statements, and deviations suppressed by sufficiently high powers in the system size should be allowed \cite{garcia2018c, Jia_2018}.  Also, these criteria are sufficient but not necessary, as there are regimes that give  NAdS$_2$/NCFT$_1$ but are  beyond the description of chord diagram combinatorics, such as the fixed $p$ and $N\to \infty$ limit of SYK, which violates the second criterion by having holonomies  that are untunable at large $N$. Nonetheless, these criteria should not be violated too violently. For example, if we  strongly violate the uniformity requirement  by assigning to each hypercube edge an independently random phase, we would end up with radically different chord combinatorics  that do not deliver NAdS$_2$/NCFT$_1$; a local spin chain model would strongly violate the second criterion by having frustrations only on a vanishingly small fraction of the graph faces.  Tentatively, the uniformity requirement will be relaxed to some smooth-variation requirement in a broader setting, but we do not have a quantitative description at present. It is even less clear to us how the second criterion should be relaxed.

Finally, since NAdS$_2$ appears as the long-throat part of a higher-dimensional geometry, we expect a large timescale separation in the dual CFT$_D$ ($D>1$), entailing an adiabatic scenario. We speculate that such random frustrations can arise as Berry curvatures when the fast degrees of freedom are integrated out \cite{mead1979b,mead1980a, berry1984}.
 \acknowledgments{
 We thank  Jacobus Verbaarschot, Antonio Garc\'\i a-Garc\'\i a, Dario Rosa and Alexander Abanov for valuable discussions. This work is supported by the ISF (2159/22), PBC (76552401) and by the DIP foundation. YJ is additionally supported by the United States–Israel Binational Science Foundation (BSF) under Grant No. 2018068, by the Minerva Foundation with funding from the Federal German Ministry for Education and Research, by the Koshland postdoctoral fellowship and by a research grant from Martin Eisenstein.
}

\bibliographystyle{apsrev4-2}
\bibliography{prlrefs.bib}

\end{document}